# Stacking fault segregation imaging with analytical field ion microscopy

Felipe F. Morgado[1], Leigh Stephenson[1], Shalini Bhatt[1], Christoph Freysoldt[1], Steffen Neumeier[2], Shyam Katnagallu[1], Aparna P.A. Subramanyam[3], Isabel Pietka[3], Thomas Hammerschmidt[3], François Vurpillot[4], and Baptiste Gault[1,5]

[1]Max-Planck-Institut für Eisenforschung, Max-Planck-Str. 1, 40237, Düsseldorf, Germany

[2]Friedrich-Alexander-Universität Erlangen-Nürnberg, Materials Science and Engineering, Institute 1, Erlangen, Germany

[3]Interdisciplinary Centre for Advanced Materials Simulation, Ruhr-Universität Bochum, Bochum, Germany

[4]Normandie Université, UNIROUEN, INSA Rouen, CNRS, Groupe de Physique des Matériaux, Rouen 76000, France

[5]Department of Materials, Royal School of Mines, Imperial College London, London, SW7 2AZ, UK

Corresponding Author: Felipe F. Morgado | Baptiste Gault <f.ferraz@mpie.de | b.gault@mpie.de>

**Abstract**

Stacking faults (SF) are important structural defects that play an essential role in the deformation of engineering alloys. However, direct observation of stacking faults at the atomic scale can be challenging. Here, we use the analytical field ion microscopy (aFIM), including density-functional theory informed contrast estimation, to image local elemental segregation at SFs in a creep-deformed solid solution single crystal alloy of Ni-2 at.% W. The segregated atoms are imaged brightly, and time-of-flight spectrometry allows for their identification as W. We also provide the first quantitative analysis of trajectory aberration, with a deviation of approximately 0.4 nm, explaining why atom probe tomography could not resolve these segregations. Atomistic simulations of substitutional W atoms at an edge dislocation in fcc Ni using an analytic bond-order potential indicate that the experimentally observed segregation is due to the energetic preference of W for the center of the stacking fault, contrasting with e.g., Re segregating to partial dislocations. Solute segregation to SF can hinder dislocation motion, increasing the strength of Ni-based superalloys. Yet direct substitution of Re by W envisaged to lower superalloys' costs, requires extra consideration in alloy design since these two solutes do not have comparable interactions with structural defects during deformation.

## Introduction

Stacking faults (SF) are a significant planar defect contributing to the mechanical properties of crystalline materials due to their influence on the activities of dislocation movements and restrictions in occurring cross slips and dislocation climbs (Zaddach et al., 2013; Hull & Bacon, 2011; Zaynullina et al., 2017). They can be formed during crystal growth, activities of partial dislocations during plastic deformation, or result from the dissociation of a full dislocation into two Shockley partial dislocations, where the distance between them is inversely proportional to the stacking fault energy (SFE) (Li et al., 2010). Low SFE in metallic materials promotes the process of deformation twinning (El-Danaf et al., 1999; Greulich & Murr, 1979; Rémy & Pineau, 1976), and the strength of materials and grain boundaries seems to increase with the density of twin boundaries (Zaynullina et al., 2017; Schneider et al., 2020). Increasing solute concentrations is one way to decrease the SFE significantly in some alloys. In these cases, the solute segregates to the SF, known as the Suzuki effect (Suzuki, 1952), imposing additional resistance to gliding dislocations, slowing down their velocity, and increasing strength (Oikawa & Iijima, 2008).

Despite all the studies regarding the effects of stacking fault and solute segregation on stacking fault on a material's properties, a direct measurement remains challenging. Nevertheless, indirect measurements of stacking faults have been carried out to better understand the underlying deformation mechanisms. For instance, stacking fault studies using electron channeling contrast imaging (ECCI) in a scanning electron microscope (SEM) equipped with a field emission gun of deformed high-alloyed transformation-induced plasticity steels (Weidner et al., 2011). Sharma et al. (2020) measured stacking faults in gold under high pressures (shock-compressed state) by X-ray diffraction (XRD). In this case, the stacking fault induces a shift and broadening in the XRD peaks. Electrical transport measurement of highly oriented pyrolitic graphite (HOPG) using the atomic force microscope by Koren et al. (2014) gives information on stacking fault density and spacing from the non-linearity of the I/V curve. Li et al. (2010) used transmission electron microscopy (TEM) to study SF and its interaction with a dislocation in a deformed polycrystalline Mg (Smith et al., 2016, 2019). Contradictory atom probe tomography (APT) studies observed higher content of Re on SF and topologically close-packed phases (TCP) (Wu et al., 2020; Xia et al., 2020), while others could not find any clustering on SF (Katnagallu et al., 2019) or matrix (Fleischmann et al., 2015).

Field ion microscopy (FIM) has a high potential to be a reliable alternative for imaging SF and looking at segregation as it is less affected by trajectory aberration than APT, where the surface atoms are imaged with atomic spatial resolution prior to field evaporation. Katnagallu et al. (2019) imaged the segregation of Re at dislocations and vacancies using FIM and measured its composition with analytical FIM (aFIM). Atomistic simulations with coarse-grained electronic-structure methods in the same work revealed that the observed segregation originates from the energetic preference of Re atoms for the partial dislocation cores instead of the stacking-fault in between. In the case of stacking faults, the imaging contrast found in FIM has been discussed before (Fortes, 1970; Herschitz et al., 1985; Howell et al., 1976; Lynch et al., 1969; Smith et al., 1968).

The imaging of stacking faults could provide new insights into the deformation mechanisms of Ni-based superalloys. Single crystal Ni-based superalloys are materials with high resistance and strength for high-temperature applications under harsh environments such as the hot section of aircraft gas turbine engines (Kommel, 2009; Murakumo et al., 2004). This strength is possible primarily due to its unique microstructure of $L1_2$ structured $\gamma'$ phase precipitates embedded in the disordered fcc $\gamma$ matrix. Re, W, and Mo are the most effective elements in Ni-based superalloys for solid solution hardening (Fleischmann et al., 2015; Giese et al., 2020). Re is the strongest among them, with a significant influence in developing a new generation of superalloys (Xia et al., 2020). It partitions strongly to the $\gamma$ matrix and hinders dislocation movement by its large atomic size, slow diffusivity, and also for decreases the SFE (Giese et al., 2020; Liu et al., 2021; Ur Rehman et al., 2017; Wöllmer et al., 2001). However, Re has several drawbacks: high density, poor oxidation resistance (Kawagishi et al., 2006), prone to forming brittle phases (Rae & Reed, 2001), and high costs (Fink et al., 2010). In this sense, it is important to understand the mechanism through which it modifies the deformation processes, to guide the search for other elements providing solid-solution hardening of the $\gamma$ phase, such as Mo and W (Fleischmann et al., 2015).

Six Ni-based superalloys with different Re, W, and Mo contents were investigated regarding their creep performance by Fleischmann et al. (2015): without considering Re clustering, at 0 K, Re is about twice as effective as W in terms of creep resistance. Regarding the two-phase $\gamma'$-strengthened alloys, the creep properties of Re-addition increase compared to W due to stronger segregation with the $\gamma$ matrix. In this case, adding 4.3 wt.% W to replace 1 wt.% Re is necessary to achieve the same creep rates. However, Viswanathan et al. (2015) reported the segregation to SF of W in the $\gamma'$ phase by using transmission electron microscopy with energy-dispersive X-ray spectroscopy (TEM-EDS) detector. SF plays a very important role in the creep properties of Ni-based superalloys due to their impact on dislocation mobility (Yang et al., 2020). The narrow channel widths of the $\gamma$ phase and the segregation of solutes to SF might promote dislocation dissociation (Unocic et al., 2008). In addition, the Suzuki effect also decreases the SFE. For example, experimentally, it has been demonstrated that the segregation of Co and Cr solutes in Ni-based superalloys decreases SFE (Achmad et al., 2018; Bobeck & Miner, 1988; Cui et al., 2012; Leverant et al., 1971). It is also important to remember that the diffusive processes contribute significantly to strengthening such alloys at high temperatures. For example, the low mobility of Re solutes decreases vacancy and dislocation mobility (Ur Rehman et al., 2017; Wu et al., 2020).

Here, we characterized a creep-deformed binary Ni-2 at.% W alloy using APT, FIM, and aFIM and performed corresponding atomistic simulations with an analytic bond-order potential (BOP) of W solute atoms near an edge dislocation in fcc-Ni. The direct analysis of superalloys by aFIM remains very complex at this stage, so we studied this model system, i.e., a creep-deformed binary alloy, to understand the fundamental mechanism. In fact, it is easier to isolate the solute contributions to the imaging process, thereby enabling the complete analysis of the aFIM data. No W clustering was observed by APT experiments, but the composition was close to its nominal composition. Meanwhile, experimental evidence of stacking fault segregation of W solute atoms was found from FIM/aFIM and explained by segregation energies obtained from BOP simulations. This segregation to stacking fault or Suzuki effect might produce a drag force on the dislocation motion and improve the strength of the material. Understanding the magnitude of such influences will contribute to the elucidation of finding better substitutes for solid-solution hardeners of Ni-based superalloys with Re additions.

## Materials and Methods

### *Materials*

A deformed solid solution single crystal alloy of Ni-2 at.% W was analyzed. The details on alloy preparation can be found in the reference Rehman et al. (2017). In short, the cylindrical specimen was submitted to a uniaxial compression creep testing machine at a temperature of 1050 °C and constant applied stress of 20 MPa until a plastic strain of about 10 % was reached within 7 h. Specimen preparation for APT/FIM was done via lift-out and annular milling procedure, following the protocol introduced by Thompson et al. (2007) using a Thermofisher Helios plasma FIB (PFIB), which uses a xenon plasma ion source.

### *Methods*

The APT and FIM/aFIM experiments were performed in a $LEAP^{TM}$ 5000 XS (Cameca Instruments). The APT measurement was conducted in laser pulsing mode with a laser pulse energy of 40 pJ, a repetition rate of 200 kHz, and a detection rate of 5 ions per 1000 pulses on average at a base temperature of 50 K. The data processing and reconstruction were done via the commercial software AP Suite 6.1, including the file conversion from .APT to .EPOS. Meanwhile, the FIM experiments were carried out in the same instrument, i.e. in digital or electronic FIM mode, with manually-controlled standing voltage, a temperature of 40 K, Ne imaging gas of purity 99.999 %, and pressure in the analysis chamber of around $3 - 8 \times 10^{-7}$ mbar. The aFIM experiments were also performed with Ne imaging gas, the temperature of 40 K, but analysis chamber pressure of around $5 - 9 \times 10^{-8}$ mbar, pulsed voltage fraction of 20 %, and pulsed rate of 65 kHz. The converted .EPOS file is processed and analyzed by using in-house MATLAB R2019a built routines, more details in Katnagallu et al. (2021).

In addition, an algorithm was developed to detect the SF segregation on the FIM images and compared it with random distributions. In this method, the $R^2$-value of each polynomial fitting of the clustered imaged atoms is extracted. If the $R^2$-value is higher than 0.9, it is assumed to be a SF segregation and it is written in white in the images. Otherwise, it is shown in red. See supplementary information for more details.

The atomistic simulations were performed with an analytic bond-order potential (BOP) that is obtained from a coarse-grained description of the electronic structure (Drautz et al., 2015; Drautz & Pettifor, 2006). The BOPs provide a robust and transparent description of the interatomic interaction at a computational effort suitable for simulations of the long-range strain fields of edge dislocations. The calculations were performed with the BOPfox software (Hammerschmidt et al., 2019) using a new parameterization for the Ni–W system that will be published elsewhere. For completeness we compiled the parameters that were used in this work in the supplemental material. For the meaning of the parameters and the according equations we refer the reader to Ref. (Subramanyam et al., 2024).

Analogous to our previous work on Re in Ni (Katnagallu et al., 2019), an a/2 [110] edge dislocation in fcc Ni is relaxed and dissociates into two a/6 <112> partials separated by a SF. In the resulting simulation cell with 3216 atoms, a W atom is introduced at different positions in the tensile and compressive layer at the glide plane. After further atomic relaxation, we determine the relative binding energy $E_B$ of W atoms in the vicinity of the dislocation with respect to a reference W atom far away from the dislocation as

$$E_B = E_{W,dislo} - E_{ref} \qquad (1)$$

The segregation preference for stacking fault or dislocation partials is then given by a comparison of the relative binding energy where $E_B > 0$ for repulsive interaction and $E_B < 0$ for attractive interaction.

DFT was performed in the plane-wave PAW formalism with the SPHInX code (Boeck et al., 2011) using the Perdew-Burke-Ernzerhof (PBE) exchange-correlation functional (Perdew et al., 1996) with D2 van-der-Waals corrections (Grimme, 2006). The Ni(012) surface was modeled in the repeated slab approach with 9 atomic layers at the theoretical lattice constant (3.46 Å) and a vacuum separation of 17.5 Å. An electric field of 40 V/nm was included via the generalized dipole correction (Freysoldt et al., 2020). Tungsten substitution at the surface was modeled in a 3×3 surface unit cell, i.e., with 9 surface atoms. For the structure optimization, an offset 2×3×1 k-point sampling was used, equivalent to a k-point spacing of 0.13 $bohr^{-1}$. For the density of states (DOS) calculation, the k-point density in the plane was doubled (4×6×1). The DFT calculations do not include finite temperature effects. We do not expect thermal effects (meV energy scale) to play a significant role in the tunneling process on the 20 eV energy scale. Certainly, thermal vibrations of the surface atoms will broaden the features, however by less than 0.1Å, which would not affect the appearance significantly.

## Results

APT experiments were conducted on the creep deformed NiW alloy to determine its composition and investigate any possibility of solute segregation to stacking faults, as shown in Figure 1. The composition found was Ni 97.3, W 1.9, along with some impurities (e.g., Si, Cr, Fe, H) 0.8 at.%, which is close to the nominal composition. Figure 1(a) is the tomographic reconstruction and Figure 1(b) shows the corresponding first-nearest neighbor W atoms distribution along with the corresponding random distribution. A comparison indicates no segregation. Figure 2 shows a FIM image of the NiW alloy formed by using the field ionization events from an aFIM experiment. Some of the crystalline poles of the face-centered cubic (fcc) crystal structure are visible and atomic resolution is achieved on atomic planes with high Miller indices.

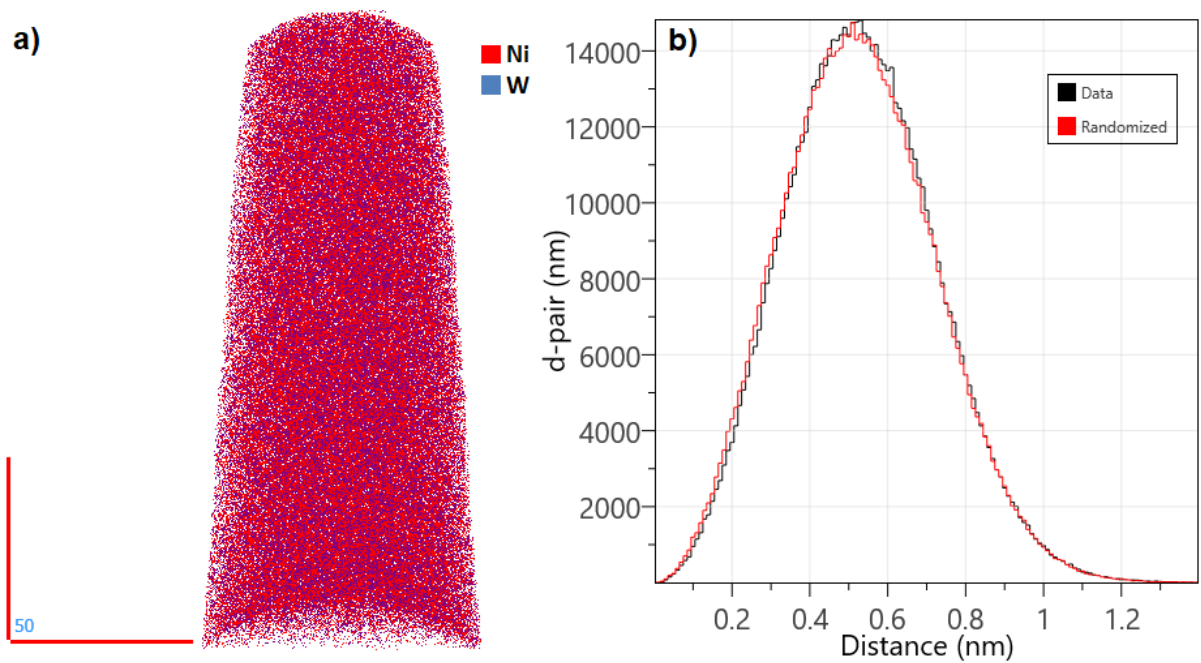

*Figure 1- Atom probe tomography laser measurement at 50 K, and laser power of 40 pJ at 200 kHz pulse rate of the crept sample. (a) Typical APT reconstruction. (b) Histogram of the distance between tungsten atoms.*

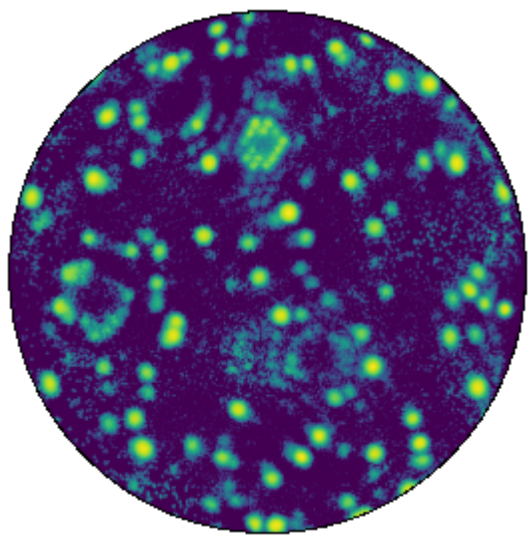

*Figure 2- FIM image created by integrating $5 \times 10^5$ successive detector hits mostly from the field ionized events in an aFIM experiment with Ne imaging gas ( $7 \times 10^{-7}$ mbar).*

### Experimental investigation of stacking fault in crept NiW alloy

Figure 3 shows typical field ion micrographs of the crept alloy with the stacking fault highlighted. The stacking faults crossing the center of the highlighted poles are observed with a few segregated brightly imaging atoms. The contrast follows the discussed geometrical models (Herschitz et al., 1985) and some of the past experimental results (Howell et al., 1976). Figure 3 also shows the $R^2$ values of the polynomial fitting described in the supplementary information. Values higher than 0.9 indicates a stacking fault.

Analytical FIM was used to identify the segregated element under $10^{-7}$ mbar of Ne imaging gas with a 20 % voltage pulsed fraction. Figure 4 shows the recreated FIM image using $10^5$ successive detector hits. The relatively lower quantity of hits compared to the quantity used to create the FIM image of Figure 2 helps reduce the noise from the field ionized events and facilitates the data analysis of the field evaporated ions through temporal and spatial correlation filtering (Morgado et al., 2023). The brightly imaged atoms at the SF are observed before the field evaporation occurs to measure their time-of-flight, i.e., in the next recreated FIM image, these atoms are no longer imaged as they have been field evaporated. Both micrographs in Figure 4(a-b) show the segregation of solute atoms to SF. The aFIM analysis gives evidence of the segregated ion being (a) $W^{2+}$ and (b) $W^{3+}$.

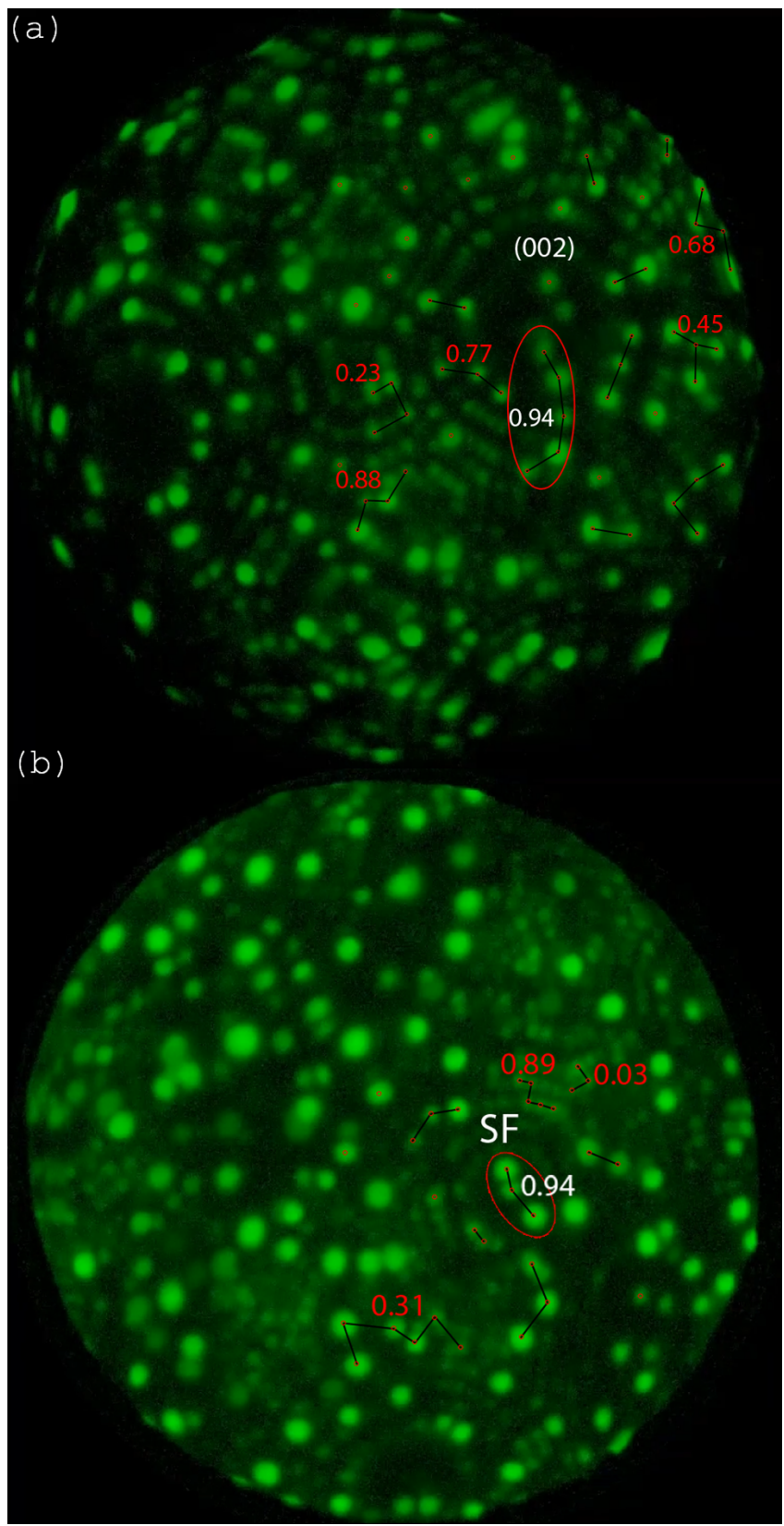

Figure 3 – Field ion micrograph of the deformed alloy with $8 \times 10^{-7}$ mbar of Ne as imaging gas. (a) W atoms are segregated to the stacking fault crossing the center of the (002) pole. (b) Three W atoms are segregated to the stacking fault observed on the highlighted pole. Identified atoms close to each other were connected and fitted with a polynomial. The $R^2$ value is shown beside the connected

atoms. White values are higher than 0.9, meaning a SF segregation was detected (highlighted in red)

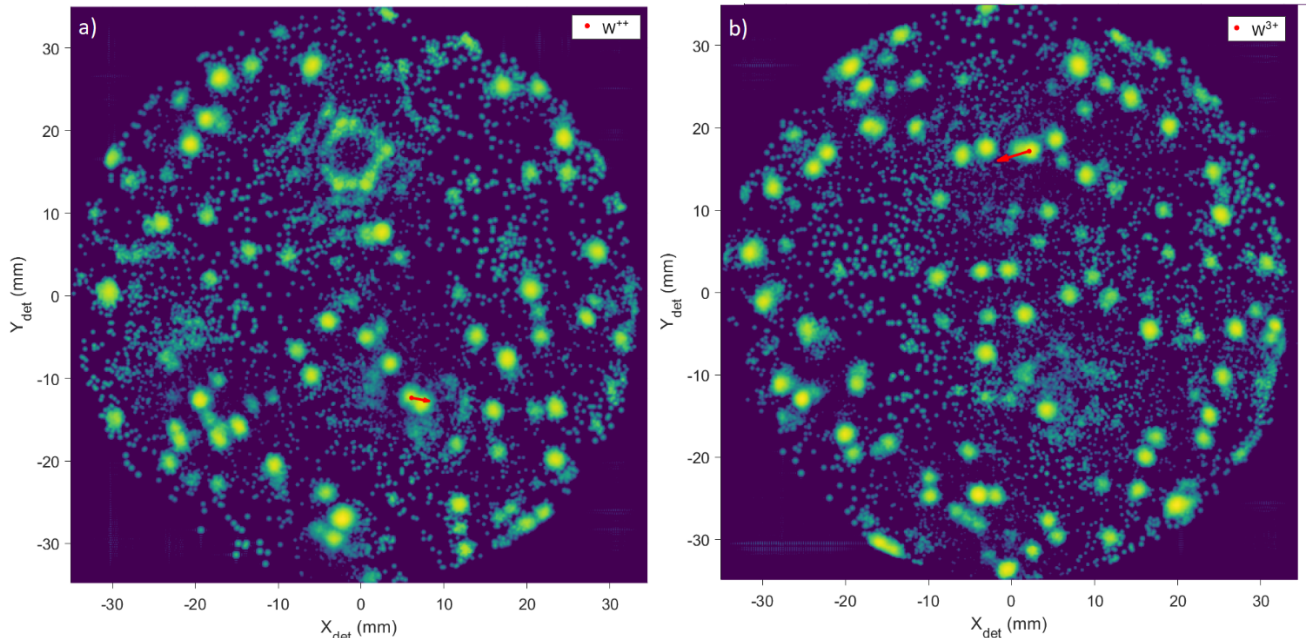

Figure 4 – Analytical FIM experiment with same conditions and specimen of Figure 2, but the image is formed with $10^5$ successive detector impact position. The red arrow indicates the distance of the atom position before and after the field evaporation. The arrow starts at the field ionized position (before) and points to the position observed on the detector (after). (a) Segregation of $W^{++}$ solute to stacking fault crossing the center of the pole. (b) Segregation of $W^{3+}$ solute to the stacking fault.

### Density functional theory for contrast

Imaging contrast in FIM is associated with the ionization probability of gas atoms near a surface under 10-100 V/nm electric fields. To better understand contrast variations among atoms observed in FIM, we employed density functional theory (DFT) in combination with the Tersoff-Hamann approximation that links electron tunneling from the gas atom to the surface. Based on our new EXTRA approach (extrapolated tail via reverse integration algorithm) to circumvent numerical noise in the tails of the wavefunctions (Bhatt et al., 2023), we demonstrate chemical brightness contrast for W in Ni(012) surface.

To evaluate the FIM contrast, we consider tunneling into empty surface states up to ~5 eV above the Fermi level via the Tersoff-Hamann approximation, i.e., from the local DOS above the surface. For each state, we consider tunneling at a height where the Ne atom's ionization level would be energetically aligned with the target state. In this case, we assume the Ne energy level lies 21.5 eV (the ionization energy of Ne) below the local electrostatic potential in the vacuum region. Thus, we ignore any shifts in the ionization energy from its gas phase value that arise from the presence of the strong field and the vicinity of the metal surface. We have verified that changes in ionization level do not modify the qualitative picture, but may moderately affect imaging contrast, see (Bhatt et al., 2023). As the electron potential increases away from the surface due to the field, the target energy thereby determines the Ne location where tunneling-induced ionization is possible. On the other hand, the wavefunctions - and hence the local DOS - decay steeply into the vacuum. This decay renders tunneling into higher-lying states increasingly unlikely, effectively imposing an upper limit of relevant energies. We found by inspection that an energy interval up to 5 eV above the Fermi level is sufficient to capture the relevant contributions to the simulated FIM intensity.

Figure 5 (a) shows the FIM simulation and confirms that bright atoms are W atoms surrounded by Ni atoms. The imaging contrast here arises from the local electronic structure rather than a local electric field enhancement (Vurpillot et al., 2001). This electronic effect can be qualitatively explained from partial DOS plots shown in Figure 5 (b): Ni contains an almost filled d-band. Tunneling into Ni d-states is possible only for a small unoccupied fraction above the Fermi level in the spin minority channel. W, however, exhibits a very significant local density of unoccupied d-orbitals above the Fermi level in both spin channels. Therefore, brightness contrast is dominated by the available unoccupied DOS. We have observed similar findings in our previous work for Ta in Ni(012) (Morgado et al., 2021), where explained qualitatively that Ta related states appear energetically at 1-3 eV above the Fermi level, while only a few Ni states in the spin minority channel are available for tunneling electrons up to 1 eV above the Fermi level. Tunneling into these orbitals will enhance the local ionization probability and hence the FIM brightness.

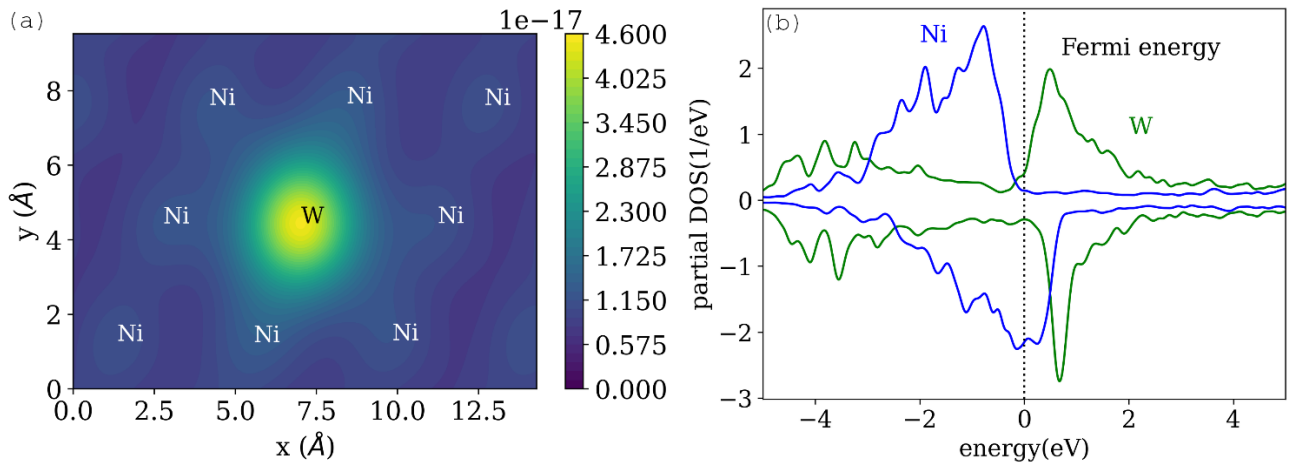

*Figure 5 - a) Simulated FIM images of W-Ni(012) for ionization energy of 21.5 eV. b) Partial density of states for Ni atom and W on the top surface layer of Ni(012).*

### *Quantification of trajectory aberrations*

A comparison of the atom position extracted from the field ion micrograph, i.e. coming from the field ionization before field evaporation, and the position of the field evaporated ion impact on the detector following filtering provides an estimation of the extent of the trajectory aberrations. Based on the local atomic environment and the difference in the evaporation field between Ni and W, it is expected to observe a similar behavior as in the case of Ni and Re reported by Katnagallu et al. (2019). Red arrows indicate this difference in the imaged (origin) and detected impact position (arrowhead), with distances of (a) 0.29 nm and (b) 0.52 nm. The magnitude of the trajectory aberration was obtained based on the distance measured on the detector and by using a point-projection for the magnification (Newman et al., 1967; Cerezo et al., 1999). The radius of curvature was deduced from the ring counting method (Tsong, 1990; Nakamura & Kuroda, 1969), see supplementary information for more details. Further data processing algorithm development is necessary to improve the accuracy of these measurements, along with sweeps through experimental parameters to optimize the retrieval of signals from the noisy data. Yet these are the first experimental measurements of the extent of trajectory aberrations in APT.

*Investigation of stacking fault in crept alloy*

The atomistic relaxation with BOP of the a/2[110] edge dislocation leads to dissociation into two a/6<112> partials separated by a SF. The SF width is consistent with the SF energy of Ni (Carter & Holmes, 1977; Shang et al., 2012). The positions of the relaxed dislocation complex of the two partials and the SF are indicated in the supercell of the simulations in Figure 6(a). Starting from this supercell, a W solute atom is inserted at different positions in the atomic layers at the SF plane. After further relaxation, the relative binding energy is computed and shown as a line scan in Figure 6(b). These are the first atomistic simulations of W near an edge dislocation in Ni.

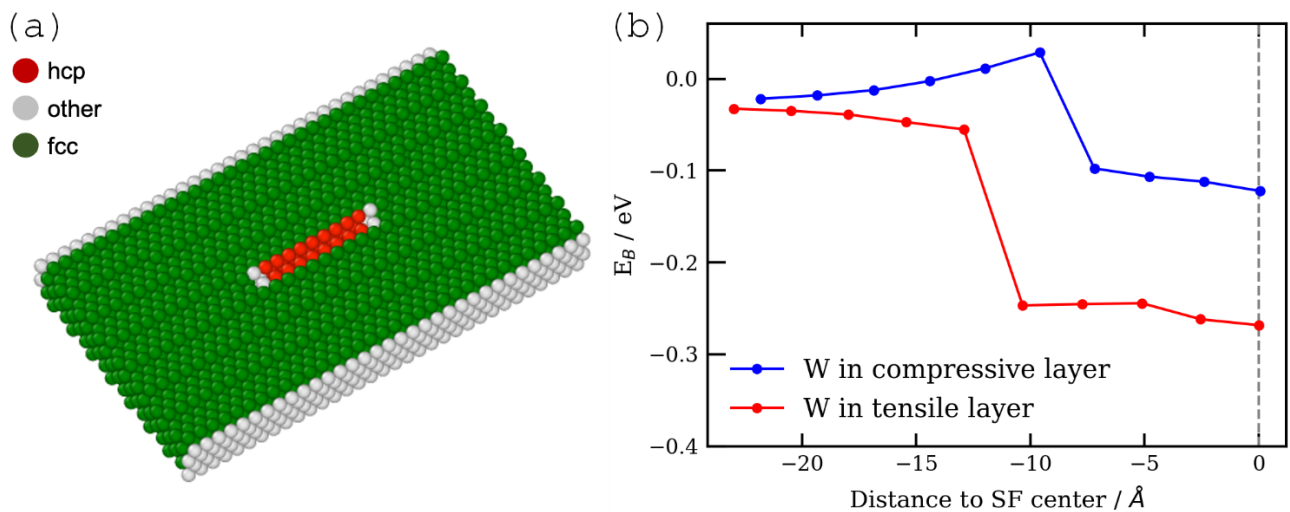

Figure 6 - Atomistic simulations of W segregation at a/2[110] edge dislocation in fcc-Ni. (a) Simulation cell of fcc Ni (green) with defect complex of two a/6<112> partials (grey) and a stacking fault (red) in between. (b) Relative binding energy of W atoms to the partials and to the SF in the layer above the SF plane (tensile) and below (compressive).

## Summary and discussion

Herein, we showcase experimental evidence of stacking fault segregation of solutes in deformed NiW alloy.

*Stacking fault observation in field ion microscopy*

As discussed earlier, direct observation of stacking faults can still be very challenging. Viswanathan et al. (2015) measured segregation in a commercial superalloy following creep deformation using TEM-EDS. They observed the solute segregation with an increase in the composition of the Cr, Co,

Mo, and W solutes in the $\gamma'$ ordered phase. Although transmission techniques can resolve atomic structures, they are still limited to the collective effects of a column or sheet of atoms through specimen thickness. On the other hand, FIM can be used to image surface atoms and 3-dimensional structural information (i.e., imaging surface atoms combined with field evaporation) (Song et al., 1996). In addition, analytical FIM or aFIM can give the chemical composition/identification of the segregated solutes in the imaged stacking faults.

The FIM specimen has a circular cross-section, curvature radius around nanometers, and atomic terraces. The specimen's topography induces an inhomogeneous field where the protruding regions have a higher field and are imaged more easily, producing a ring-shaped image for each atomic layer. Meanwhile, based on purely geometrical considerations, stacking fault defects produce atomic steps crossing the crystalline pole and their surface atoms crossing the pole would be visible. Experimentally, Howell et al. (1976) investigated the segregation of solutes to a stacking fault, mainly the segregation of niobium in a cobalt-niobium alloy. The stacking fault crosses the center of the ($\bar{1}00$) pole, and solute niobium atoms are retained. Similar results were reported by Song et al. (1996) on iridium specimens.

Here is reported a similar contrast, the segregation of solute in stacking faults imaged in the micrographs of Figure 3 and Figure 4. The FIM images show bigger and brighter atoms crossing the crystalline poles. Their size and brightness indicate that the surface atoms are W, like what was observed in past experiments (Katnagallu et al., 2019; Morgado et al., 2021). In addition, aFIM experiment identified these atoms as $W^{++}$ and $W^{3+}$ with trajectory aberration of around $\pm 0.4$ nm. The trajectory aberration was obtained using point-projection for the magnification (Newman et al., 1967; Cerezo et al., 1999) and ring counting method to obtain the radius of curvature (Tsong, 1990; Nakamura & Kuroda, 1969), see supplementary information for more details. Meanwhile, no evidence of segregation was found in the APT measurement. It is difficult to estimate the total volumetric density of defects in 3D, even with other techniques such as transmission electron microscopy or electron channelling contrast imaging in the scanning electron microscope (Zaefferer & Elhami, 2014). With the material creep deformed at high temperatures until fracture, a high defect density is expected. FIM imaging of this material systematically revealed one or more SF in the field of view accessible on the LEAP 5000 XS at our disposal. The same instrument is used for APT and FIM, so it is extremely unlikely that no SF is present in APT mode, especially considering that analyses were performed over hundreds of nanometers in depth.".

Please note that the aFIM acquisition was performed using the voltage-pulsing mode, whereas the APT experiment via laser-pulsing mode. This is because of the relatively low yield and small APT datasets obtained in voltage pulsing mode, in combination with an issue of the mass-to-charge overlap between $Ni^+$ and $W^{+++}$.

### *Stacking fault detection method*

The stacking fault detection method gives evidence that the distribution of the surface atoms in the stacking fault is not random. The algorithm is described in Figure S1, which consists of connecting nearby surface atoms via the nearest neighbors search algorithm with a distance threshold defined for the specific experiment/specimen. Each of these fragments are fitted with a polynomial of different order depending on the number of connected atoms. Linear fitting for 3 atoms, polynomial curve fitting for 4 atoms (2$^{nd}$ order), and 5-7 atoms (4$^{th}$ order). This methodology gave better results for the SF segregation studies on FIM experiments due to the different magnifications on the FIM images making the SF curved. The $R^2$ values are extracted from the fitting and the SF is identified when the values are higher than 0.9 (written in white in the images). The implemented algorithm showed consistent and successful results for the analyzed cases:

simulated pure metal with BCC crystal structure, simulated highly deformed alloy Ni-2 at.% Re, and the experimental results (FIM, aFIM) of Ni-2 at.% W. Lastly, the identified atoms are redistributed using pseudorandom values from a discrete uniform distribution. Figure S6(a) shows that the random distribution has clusters with fewer atoms compared to the SF segregation images. In addition, Figure S6(b) shows that the random distribution contains mostly low values of $R^2$ compared to the SF analyses. In results, it shows that the surface atoms found crossing the pole are not random and it is a SF segregation.

### *Segregation energy from atomistic simulations*

The atomistic simulations with BOPs confirm the experimental observation of W segregation to the SF. In similarity to Re, we find that inserted W solute atoms show an energetic preference for the SF layer under tension over the SF layer under compression (Katnagallu et al., 2019). In contrast to Re, however, the overall energetic preference of W atoms is the center of the SF and not the core of the partial dislocation. This qualitatively different segregation tendency for W and Re atoms with very similar atomic size contradicts the expectation that segregation to the dislocation complex is predominantly determined by the relation between atomic size and available local volume. Instead, further contributions like the local electronic structure may need to be taken into account.

## Conclusion

In summary, a deformed solid solution single crystal alloy of Ni-2 at.% W was characterized using APT and FIM/aFIM. No W clustering was observed in the APT analysis, and the measured composition was close to the nominal composition. Meanwhile, FIM experiments gave clear contrast of stacking faults crossing the center of the poles with the segregation of brighter atoms. aFIM experiments indicated that the brighter atoms were W ions segregating to the stacking faults, known as the Suzuki effect. The experimental findings are in very good agreement with atomistic simulations that show a preference for W solute atoms for the center of the SF in the tensile layer. In addition, differences in the detector position of the field ionization and field evaporation events give information about the trajectory aberration of each segregated solute. The brighter contrast of W was also confirmed by DFT calculations, which is further enhanced by unoccupied W d-orbitals above the Fermi level. Nevertheless, the SF segregation found in the deformed binary alloy gives good prospects for the replacement of Re solute by W, as already pursued by some research groups (Horst et al., 2020). The segregation could increase the strength of the Ni-based superalloys due to hindering the dislocation motion caused by the difference in the crystal structure. Moreover, the resulting lower stacking fault energy, its low diffusion coefficient, and lower cost than the Re element make W a promising substitute.

**Acknowledgements** This work was supported by a scholarship from The International Max Planck Research School for Interface Controlled Materials for Energy Conversion (IMPRS-SurMat) for F.F.M. We appreciate the technical support of U. Tezins, C. Bross and A. Sturm at the APT/FIB facilities at Max-Planck-Institut für Eisenforschung. LTS & BG author acknowledges financial support from the ERC CoG-SHINE-771,602. The authors acknowledge the German National Science Foundation (DFG) support within the SPP-1594 (DE796/9-2, RA 659/28-2, and SCHN 735/35-2) and project C1 and A6 of the collaborative research center SFB/TR 103 (DFG project number 190389738). TH gratefully acknowledges funding of this project by computing time provided by the Paderborn Center for Parallel Computing (PC2). SB gratefully acknowledges financial support from the International Max Planck Research School for Sustainable Metallurgy (IMPRS SusMet).

Competing interests: the author(s) declare none.

## References

ACHMAD, T. L., FU, W., CHEN, H., ZHANG, C. & YANG, Z.-G. (2018). Effect of solute segregation on the intrinsic stacking fault energy of Co-based binary alloys: A first-principles study. *Journal of Alloys and*

  *Compounds* **748**, 328–337.

BHATT, S., KATNAGALLU, S., NEUGEBAUER, J. & FREYSOLDT, C. (2023). Accurate computation of chemical contrast in field ion microscopy. *Physical Review B* **107**, 235413.

BOBECK, G. E. & MINER, R. V. (1988). Effects of cobalt concentration on the relative resistance to octahedral and cube slip in nickel-base superalloys. *Metallurgical Transactions A* **19**, 2733–2739.

BOECK, S., FREYSOLDT, C., DICK, A., ISMER, L. & NEUGEBAUER, J. (2011). The object-oriented DFT program library S/PHI/nX. *Comp. Phys. Comm.* **182**, 543–554.

CARTER, C. B. & HOLMES, S. M. (1977). The stacking-fault energy of nickel. *The Philosophical Magazine: A Journal of Theoretical Experimental and Applied Physics* **35**, 1161–1172.

CEREZO, A., WARREN, P. J. & SMITH, G. D. W. (1999). Some aspects of image projection in the field-ion microscope. *Ultramicroscopy* **79**, 251–257.

CUI, C., TIAN, C., ZHOU, Y., JIN, T. & SUN, X. (2012). Dynamic strain aging in Ni base alloys with different stacking fault energy. *CD room Superalloy* E1.

DRAUTZ, R., HAMMERSCHMIDT, T., ČÁK, M. & PETTIFOR, D. G. (2015). Bond-order potentials: derivation and parameterization for refractory elements. *Modelling and Simulation in Materials Science and Engineering* **23**, 074004.

DRAUTZ, R. & PETTIFOR, D. G. (2006). Valence-dependent analytic bond-order potential for transition metals. *Physical Review B* **74**, 174117.

EL-DANAF, E., KALIDINDI, S. R. & DOHERTY, R. D. (1999). Influence of grain size and stacking-fault energy on deformation twinning in fcc metals. *Metallurgical and Materials Transactions A* **30**, 1223–1233.

FINK, P. J., MILLER, J. L. & KONITZER, D. G. (2010). Rhenium reduction—alloy design using an economically strategic element. *JOM* **62**, 55–57.

FLEISCHMANN, E., MILLER, M. K., AFFELDT, E. & GLATZEL, U. (2015). Quantitative experimental determination of the solid solution hardening potential of rhenium, tungsten and molybdenum in single-crystal nickel-based superalloys. *Acta Materialia* **87**, 350–356.

FORTES, M. A. (1970). Characterization of stacking faults with the field-ion microscope. *Philosophical Magazine* **22**, 317–327.

FREYSOLDT, C., MISHRA, A., ASHTON, M. & NEUGEBAUER, J. (2020). Generalized dipole correction for charged surfaces in the repeated-slab approach. *Physical Review B* **102**, 045403.

GIESE, S., BEZOLD, A., PRÖBSTLE, M., HECKL, A., NEUMEIER, S. & GÖKEN, M. (2020). The Importance of Diffusivity and Partitioning Behavior of Solid Solution Strengthening Elements for the High Temperature Creep Strength of Ni-Base Superalloys. *Metallurgical and Materials Transactions A* **51**, 6195–6206.

GREULICH, F. & MURR, L. E. (1979). Effect of Grain size, dislocation cell size and deformation twin spacing on the residual strengthening of shock-loaded nickel. *Materials Science and Engineering* **39**, 81–93.

GRIMME, S. (2006). Semiempirical GGA-type density functional constructed with a long-range dispersion correction. *Journal of Computational Chemistry* **27**, 1787–1799.

HAMMERSCHMIDT, T., SEISER, B., FORD, M. E., LADINES, A. N., SCHREIBER, S., WANG, N., JENKE, J., LYSOGORSKIY, Y., TEIJEIRO, C., MROVEC, M., CAK, M., MARGINE, E. R., PETTIFOR, D. G. & DRAUTZ, R. (2019). BOPfox program for tight-binding and analytic bond-order potential calculations. *Computer Physics Communications* **235**, 221–233.

HERSCHITZ, R., SEIDMAN, D. N. & BROKMAN, A. (1985). Solute-atom segregation and two-dimensional phase transitions in stacking faults: an atom-probe field-ion microscope study. *Le Journal de Physique Colloques* **46**, C4-451-C4-464.

HORST, O. M., ADLER, D., GIT, P., WANG, H., STREITBERGER, J., HOLTKAMP, M., JÖNS, N., SINGER, R. F., KÖRNER, C. & EGGELER, G. (2020). Exploring the fundamentals of Ni-based superalloy single crystal (SX) alloy design: Chemical composition vs. microstructure. *Materials & Design* **195**, 108976.

HOWELL, P. R., FLEET, D. E., HILDON, A. & RALPH, B. (1976). Field-ion microscopy of segregation to planar imperfections. *Journal of Microscopy* **107**, 155–167.

HULL, D. & BACON, D. J. (2011). Chapter 1 - Defects in Crystals. In *Introduction to Dislocations (Fifth Edition)*, Hull, D. & Bacon, D. J. (Eds.), pp. 1–20. Oxford: Butterworth-Heinemann https://www.sciencedirect.com/science/article/pii/B9780080966724000013.

KATNAGALLU, S., MORGADO, F. F., MOUTON, I., GAULT, B. & STEPHENSON, L. T. (2021). Three-Dimensional Atomically Resolved Analytical Imaging with a Field Ion Microscope. *Microscopy and Microanalysis* 1–16.

KATNAGALLU, S., STEPHENSON, L. T., MOUTON, I., FREYSOLDT, C., SUBRAMANYAM, A. P. A., JENKE, J., LADINES, A. N., NEUMEIER, S., HAMMERSCHMIDT, T., DRAUTZ, R., NEUGEBAUER, J., VURPILLOT, F., RAABE, D. & GAULT, B. (2019). Imaging individual solute atoms at crystalline imperfections in metals. *New Journal of Physics* **21**, 123020.

KAWAGISHI, K., HARADA, H., SATO, AKIHIRO, SATO, ATSUSHI & KOBAYASHI, T. (2006). The oxidation properties of fourth generation single-crystal nickel-based superalloys. *JOM* **58**, 43–46.

KOMMEL, L. (2009). Viscoelastic Behavior of a Single-Crystal Nickel-Base Superalloy. *Medziagotyra* **15**.

KOREN, E., KNOLL, A. W., LÖRTSCHER, E. & DUERIG, U. (2014). Direct experimental observation of stacking fault scattering in highly oriented pyrolytic graphite meso-structures. *Nat. Commun.* **5**, 5837.

LEVERANT, G. R., KEAR, B. H. & OBLAK, J. M. (1971). The influence of matrix stacking fault energy on creep deformation modes in Γ' precipitation-hardened nickel-base alloys. **2**, 2305–2306.

LI, B., YAN, P. F., SUI, M. L. & MA, E. (2010). Transmission electron microscopy study of stacking faults and their interaction with pyramidal dislocations in deformed Mg. *Acta Materialia* **58**, 173–179.

LIU, C., ZHANG, X., WANG, C., YU, T., ZHANG, Y., LI, H. & ZHANG, Z. (2021). The micro-mechanism of rhenium promoting the formation of stacking faults in the Ni-based model single crystal superalloys. *Journal of Alloys and Compounds* **851**, 156777.

LYNCH, J. T., CRANSTOUN, G. K. L., SMITH, D. A. & SMITH, G. D. W. (1969). Field-ion Microscopic Observation of a Stacking Fault in Pure Iron. *Nature* **222**, 637–639.

MORGADO, F. F., KATNAGALLU, S., FREYSOLDT, C., KLAES, B., VURPILLOT, F., NEUGEBAUER, J., RAABE, D., NEUMEIER, S., GAULT, B. & STEPHENSON, L. T. (2021). Revealing atomic-scale vacancy-solute interaction in nickel.

*Scripta Materialia* **203**, 114036.

MORGADO, F. F., STEPHENSON, L., ROUSSEAU, L., VURPILLOT, F., EVERTZ, S., SCHNEIDER, J. M. & GAULT, B. (2023). Improving Spatial and Elemental Associations in Analytical Field Ion Microscopy. *Microscopy and Microanalysis* ozad039.

MURAKUMO, T., KOBAYASHI, T., KOIZUMI, Y. & HARADA, H. (2004). Creep behaviour of Ni-base single-crystal superalloys with various γ′ volume fraction. *Acta Materialia* **52**, 3737–3744.

NAKAMURA, S. & KURODA, T. (1969). On field-evaporation end forms of a bcc metal surface observed by a field ion microscope. *Surface Science* **17**, 346–358.

NEWMAN, R. W., SANWALD, R. C. & HREN, J. J. (1967). A method for indexing field ion micrographs. *Journal of Scientific Instruments* **44**, 828–830.

OIKAWA, H. & IIJIMA, Y. (2008). 7 - Diffusion behaviour of creep-resistant steels. In *Creep-Resistant Steels*, *Woodhead Publishing Series in Metals and Surface Engineering*, Abe, F., Kern, T.-U. & Viswanathan, R. (Eds.), pp. 241–264. Woodhead Publishing https://www.sciencedirect.com/science/article/pii/B9781845691783500073.

PERDEW, J. P., BURKE, K. & ERNZERHOF, M. (1996). Generalized Gradient Approximation Made Simple. *Physical Review Letters* **77**, 3865–3868.

RAE, C. M. F. & REED, R. C. (2001). The precipitation of topologically close-packed phases in rhenium-containing superalloys. *Acta Materialia* **49**, 4113–4125.

RÉMY, L. & PINEAU, A. (1976). Twinning and strain-induced f.c.c. - h.c.p. transformation on the mechanical properties of Co-Ni-Cr-Mo alloys. *Materials Science and Engineering* **26**, 123–132.

SCHNEIDER, M., GEORGE, E. P., MANESCAU, T. J., ZÁLEŽÁK, T., HUNFELD, J., DLOUHÝ, A., EGGELER, G. & LAPLANCHE, G. (2020). Analysis of strengthening due to grain boundaries and annealing twin boundaries in the CrCoNi medium-entropy alloy. *International Journal of Plasticity* **124**, 155–169.

SHANG, S. L., ZACHERL, C. L., FANG, H. Z., WANG, Y., DU, Y. & LIU, Z. K. (2012). Effects of alloying element and temperature on the stacking fault energies of dilute Ni-base superalloys. *Journal of Physics: Condensed Matter* **24**, 505403.

SHARMA, S. M., TURNEAURE, S. J., WINEY, J. M., RIGG, P. A., SINCLAIR, N., WANG, X., TOYODA, Y. & GUPTA, Y. M. (2020). Real-Time Observation of Stacking Faults in Gold Shock Compressed to 150 GPa. *Phys. Rev. X* **10**, 011010.

SMITH, D. A., FORTES, M. A., KELLY, A. & RALPH, B. (1968). Contrast from stacking faults and partial dislocations in the field-ion microscope. *The Philosophical Magazine: A Journal of Theoretical Experimental and Applied Physics* **17**, 1065–1077.

SMITH, T. M., GOOD, B. S., GABB, T. P., ESSER, B. D., EGAN, A. J., EVANS, L. J., MCCOMB, D. W. & MILLS, M. J. (2019). Effect of Stacking Fault Segregation and Local Phase Transformations on Creep Strength in Ni-Base Superalloys. https://papers.ssrn.com/abstract=3361809 (Accessed February 21, 2023).

SMITH, T. M., UNOCIC, R. R., DEUTCHMAN, H. & MILLS, M. J. (2016). Creep deformation mechanism mapping in nickel base disk superalloys. *Materials at High Temperatures* **33**, 372–383.

SONG, S. G., CHEN, C. L. & TSONG, T. T. (1996). Field ion microscopy observation of intrinsic stacking faults in iridium. *Materials Science and Engineering: A* **212**, 119–122.

SUBRAMANYAM, A. P. A., JENKE, J., LADINES, A. N., DRAUTZ, R. & HAMMERSCHMIDT, T. (2024). Parametrization protocol and refinement strategies for accurate and transferable analytic bond-order potentials: Application to Re. *Physical Review Materials* **8**, 013803.

SUZUKI, H. (1952). Chemical interaction of solute atoms with dislocations. *Sci. Rep. Res. Inst. Tohoku Univ. A* **4**, 455–463.

THOMPSON, K., LAWRENCE, D., LARSON, D. J., OLSON, J. D., KELLY, T. F. & GORMAN, B. (2007). In situ site-specific specimen preparation for atom probe tomography. *Ultramicroscopy* **107**, 131–139.

TSONG, T. T. (1990). *Atom-Probe Field Ion Microscopy*. Cambridge University Press https://doi.org/10.1017/cbo9780511599842.

UNOCIC, R., KOVARIK, L., SHEN, C., SAROSI, P., WANG, Y., LI, J., GHOSH, S. & MILLS, M. (2008). Deformation Mechanisms in Ni-Base Disk Superalloys at Higher Temperatures. *Proceedings of the International Symposium on Superalloys*.

UR REHMAN, H., DURST, K., NEUMEIER, S., SATO, A., REED, R. & GÖKEN, M. (2017). On the temperature dependent strengthening of nickel by transition metal solutes. *Acta Materialia* **137**, 54–63.

VISWANATHAN, G. B., SHI, R., GENC, A., VORONTSOV, V. A., KOVARIK, L., RAE, C. M. F. & MILLS, M. J. (2015). Segregation at stacking faults within the γ' phase of two Ni-base superalloys following intermediate temperature creep. *Scripta Materialia* **94**, 5–8.

VURPILLOT, F., BOSTEL, A. & BLAVETTE, D. (2001). A new approach to the interpretation of atom probe field-ion microscopy images. *Ultramicroscopy* **89**, 137–144.

WEIDNER, A., GLAGE, A., SPERLING, L. & BIERMANN, H. (2011). Observation of stacking faults in a scanning electron microscope by electron channelling contrast imaging. *International Journal of Materials Research* **102**, 3–5.

WÖLLMER, S., MACK, T. & GLATZEL, U. (2001). Influence of tungsten and rhenium concentration on creep properties of a second generation superalloy. *Materials Science and Engineering: A* **319–321**, 792–795.

WU, X., MAKINENI, S. K., LIEBSCHER, C. H., DEHM, G., MIANROODI, J. R., SHANTHRAJ, P., SVENDSEN, B., BÜRGER, D., EGGELER, G., RAABE, D. & GAULT, B. (2020). Unveiling the Re effect in Ni-based single crystal superalloys. *Nature Communications* **11**.

XIA, W., ZHAO, X., YUE, L. & ZHANG, Z. (2020). A review of composition evolution in Ni-based single crystal superalloys. *Journal of Materials Science & Technology* **44**, 76–95.

YANG, W., QU, P., SUN, J., YUE, Q., SU, H., ZHANG, J. & LIU, L. (2020). Effect of alloying elements on stacking fault energies of γ and γ' phases in Ni-based superalloy calculated by first principles. *Vacuum* **181**, 109682.

ZADDACH, A. J., NIU, C., KOCH, C. C. & IRVING, D. L. (2013). Mechanical Properties and Stacking Fault Energies of NiFeCrCoMn High-Entropy Alloy. *JOM* **65**, 1780–1789.

Zaefferer, S. & Elhami, N.-N. (2014). Theory and application of electron channelling contrast imaging under controlled diffraction conditions. *Acta Materialia* **75**, 20–50.

Zaynullina, L., Alexandrov, I. & Wei, W. (2017). Effect of the stacking fault energy on the mechanical properties of pure Cu and Cu-Al alloys subjected to severe plastic deformation Bratan, S., Gorbatyuk, S., Leonov, S. & Roshchupkin, S. (Eds.). *MATEC Web of Conferences* **129**, 02032.

# Stacking fault segregation imaging with analytical field ion microscopy

Felipe F. Morgado[1], Leigh Stephenson[1], Shalini Bhatt[1], Christoph Freysoldt[1], Steffen Neumeier[2], Shyam Katnagallu[1], Aparna P.A. Subramanyam[3], Isabel Pietka[3], Thomas Hammerschmidt[3], François Vurpillot[4], and Baptiste Gault[1,5]

[1]*Max-Planck-Institut für Eisenforschung, Max-Planck-Str. 1, 40237, Düsseldorf, Germany*

[2]*Materials Science and Engineering, Institute 1, Friedrich-Alexander-Universität Erlangen-Nürnberg, Erlangen, Germany*

[3]*Interdiscuplinary Centre for Advanced Materials Simulation, Ruhr-UniversitätBochum, Bochum, Germany*

[4]*Normandie Université, UNIROUEN, INSA Rouen, CNRS, Groupe de Physique des Matériaux, Rouen 76000, France*

[5]*Department of Materials, Royal School of Mines, Imperial College London, London, SW7 2AZ, UK*

Corresponding Author: Felipe F. Morgado | Baptiste Gault <f.ferraz@mpie.de | b.gault@mpie.de>

*Stacking fault segregation detection method*

A general description of the method is shown in Figure S7. The algorithm is based on image analysis and it could be applied to similar images, i.e., it is not restricted to aFIM experiments. In summary, the image is read and cropped in the region of interest (a specific pole), then the imaged surface atoms are identified based on their brightness via peak-finding methods (Natan, 2021). The identified atom positions are subjected to a cluster-finding algorithm using 2nd- or 3rd-nearest neighbors with a distance threshold (55 for NiW experiments and 15 otherwise), which depends on the magnification and hence on the considered image. Lastly, a line joining the position of these clustered atoms is defined by fitting a polynomial with different orders (e.g., 1, 2, or 4) with different constraints as shown in Figure S7, depending on the number of connected atoms. The fit is performed with the coordinates of the identified atoms in the image, where it can be rotated depending on how the feature is imaged. Although segregation to stacking faults might be expected to appear as line-like features, the nature of the projection of the highly curved surface of the specimen onto the screen in FIM will cause them to appear as curved lines. As a consequence, they cannot be fitted with a straight line because their curvature will depend on the magnification and probably the location of the feature within the field of view. In this way, we opted for a versatile function to describe a non-linear feature, namely a polynomial function. This ad-hoc approach was tested on experimental and simulated field ion images and provided the necessary information for this particular study. This fitting is used to extract the $R^2$-value of each fitting, and if the $R^2$-value is higher than 0.9, it is assumed to be a SF segregation. The $R^2$ values are written in the analyzed image in red for those lower than 0.9 and in white for those higher than 0.9. Values lower than 0.02 are not shown in the images. The order of the polynomial fitting depends on how many atoms are connected because of the nature of the possible segregation and the expected localization of the surface atoms. Anything less than 3 atoms are not considered for the algorithm.

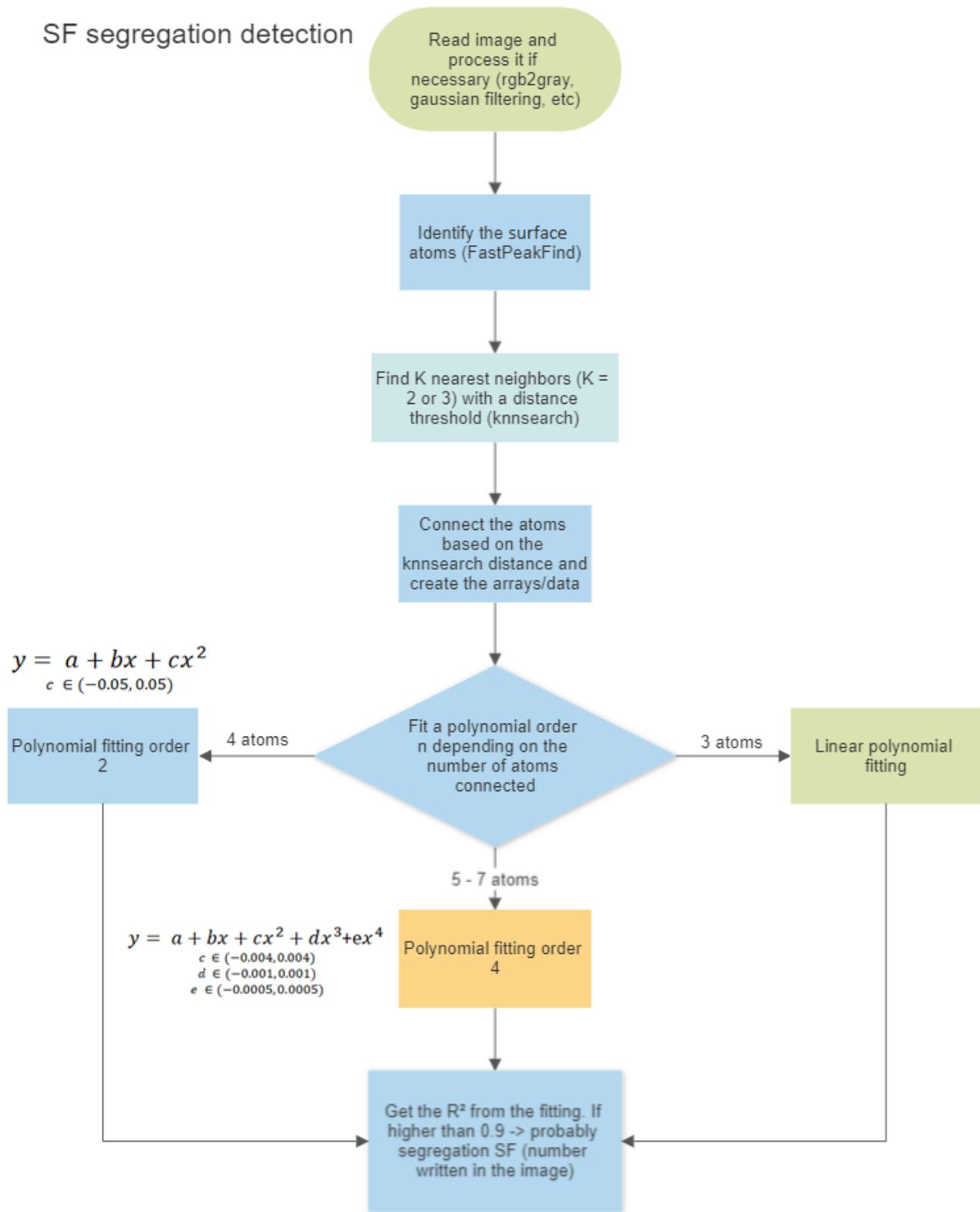

*Figure S7 - Flowchart for the detection of the stacking fault segregation in FIM-like images.*

The FIM micrograph of pure metal with a body-centered crystal structure with a radius of 12

nm was simulated using the protocol discussed in Katnagallu et al. (2019). Figure S8(a) shows the simulated FIM micrograph centered on the (011) terrace. Figure S8(b) shows the surface evolution of 4 different images of the pole (113) over the simulated field evaporation process. Atoms close together are connected, forming clusters and fitted with a polynomial of different orders. The $R^2$ values are shown beside the connected atoms and are consistently below 0.83, which implies that no segregation to SF is detected. This is the correct assumption since the sample was not deformed.

In contrast, Katnagallu et al. (2019) reported simulations for a highly deformed Ni sample under uniaxial tension using a large-scale atomic/molecular massively parallel simulator (LAMMPS) software package. Artificially, up to 25 at.% of Re was then added to the formed SFs in the simulated Ni crystal. Figure S9(a) shows corresponding simulated FIM micrographs. There are many small dislocations, mostly Shockley type (1/6 <112>). The surface evolution is shown in Figure S9(b) for 4 images, where 25 atoms are field evaporated for each image. For image number 1, a SF is visible and detected since the $R^2$ value of the polynomial fit of the cluster is approximately 1.00

The clustered atoms from the simulations and the experimental results are redistributed using pseudorandom values from a discrete uniform distribution to compare with the segregated atoms. Figure S11 shows this analysis of the clustered atoms compared with the random distribution. The random distribution clusters are limited to smaller sizes, from 2 to 4 atoms (Figure S11(a)), and also a lower amount of $R^2$ values close to 1 for the polynomial fitting (Figure S11(b)). Meanwhile, the segregated clusters tend to have a higher number of atoms, from 2 to 6, and a higher amount of fittings with $R^2$ close to 1 for the polynomial fitting.

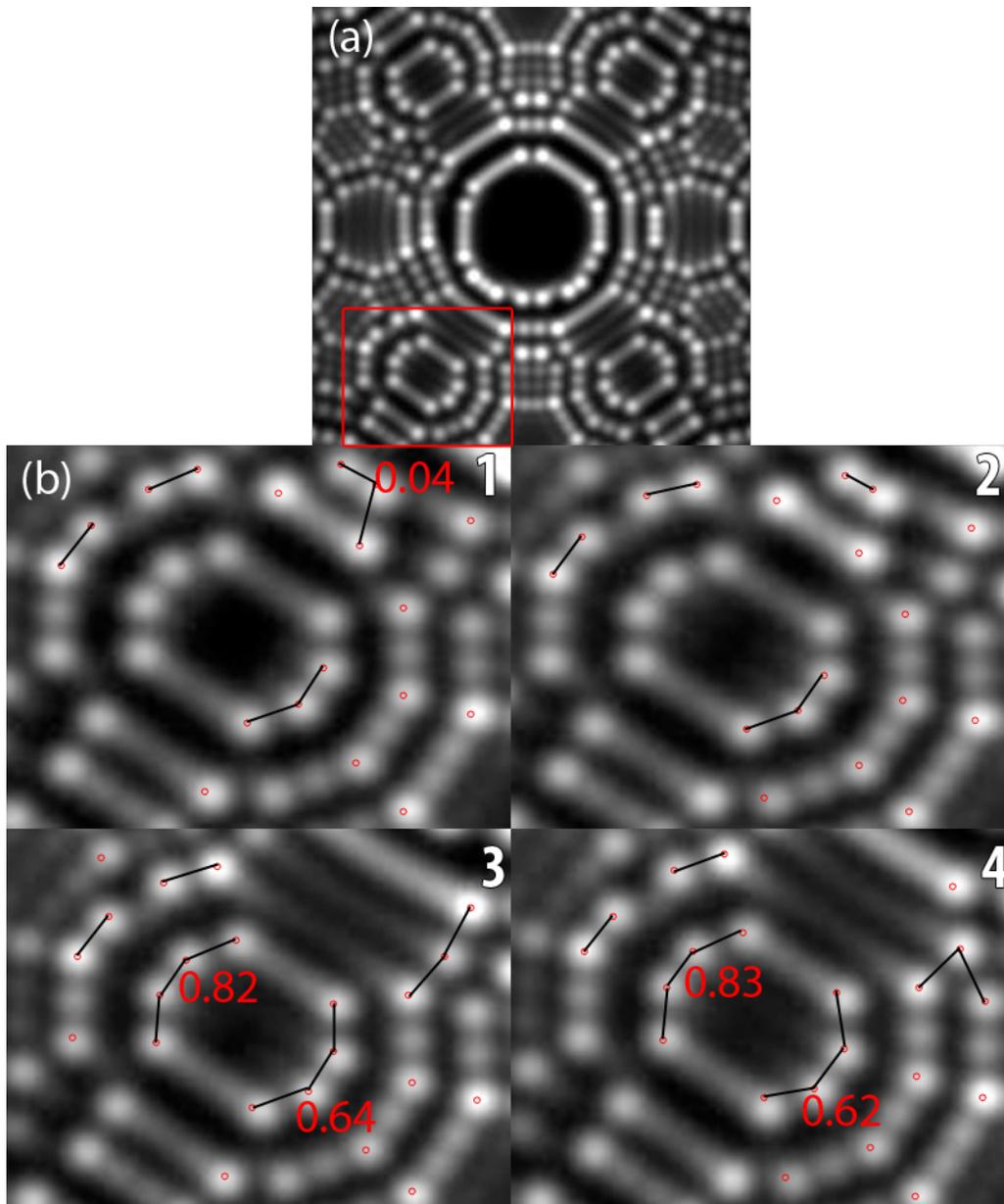

*Figure S8 – (a) Simulated FIM micrograph of pure metal with BCC crystal structure centered on the (011) terrace with individual atoms visible. Pole (113) is highlighted in red. (b) The surface evolution of 4 images by field evaporation of the simulated simple cubic specimen focused on the pole (113). Identified atoms close to each other were connected and fitted with a polynomial. The $R^2$ value is shown beside the connected atoms.*

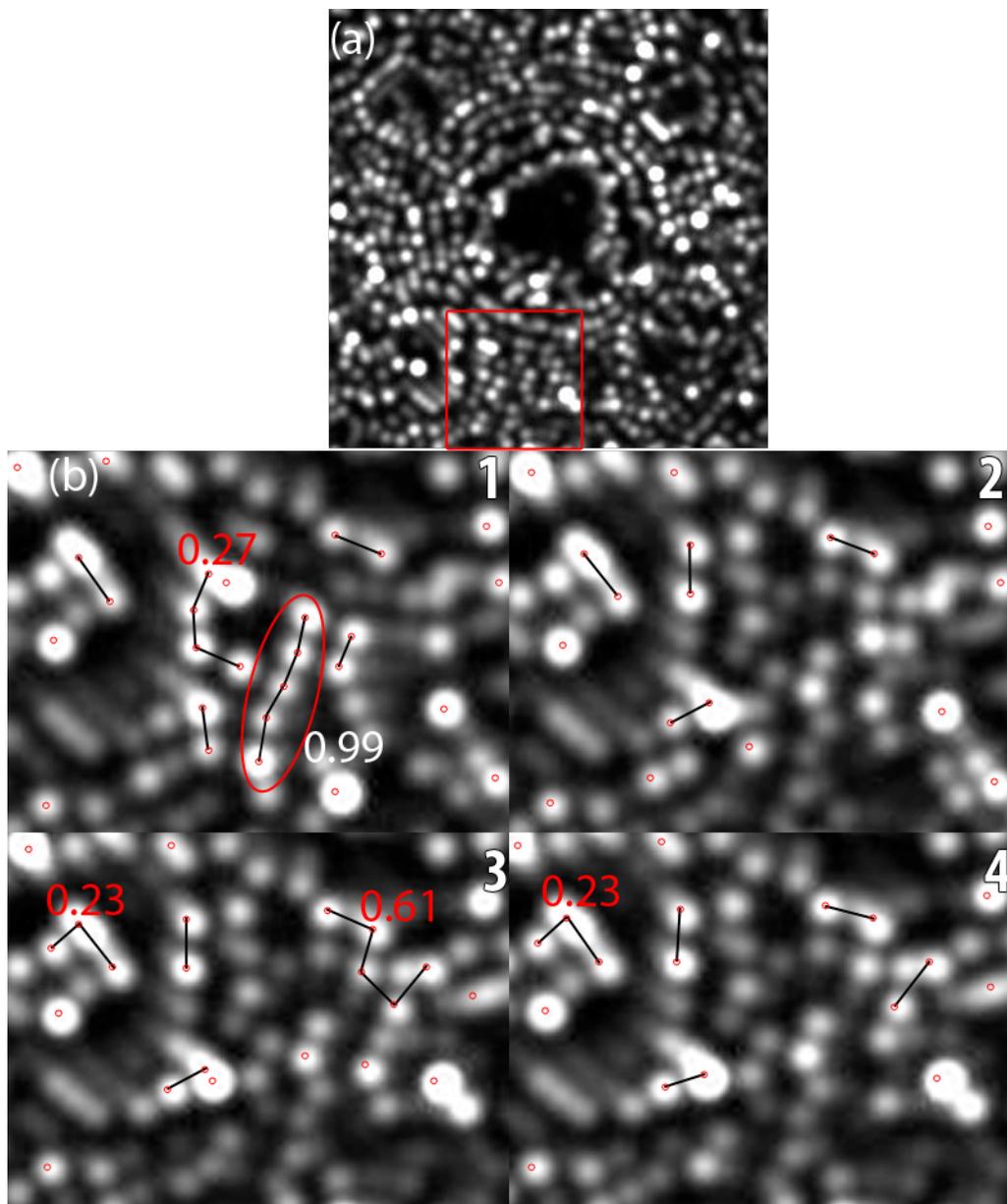

*Figure S9 – (a) Simulated FIM micrograph by Katnagallu and Vurpillot of a highly deformed alloy Ni-2 at.% Re using large-scale atomic/molecular massively parallel simulator (LAMMPS) software package (Katnagallu et al. 2019). (b) Surface evolution of 4 images by field evaporation of the deformed alloy with a focus on the highlighted area.*

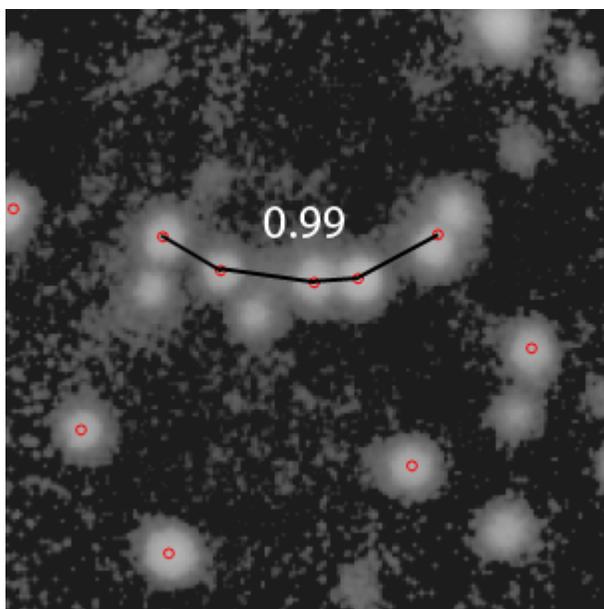

*Figure S10 – Analytical FIM micrograph in a gray scale of the SF region of Figure 5. Atoms close together are connected by lines, and the $R^2$ values of the polynomial fitting are written. White values are higher than 0.9, meaning a SF segregation was detected.*

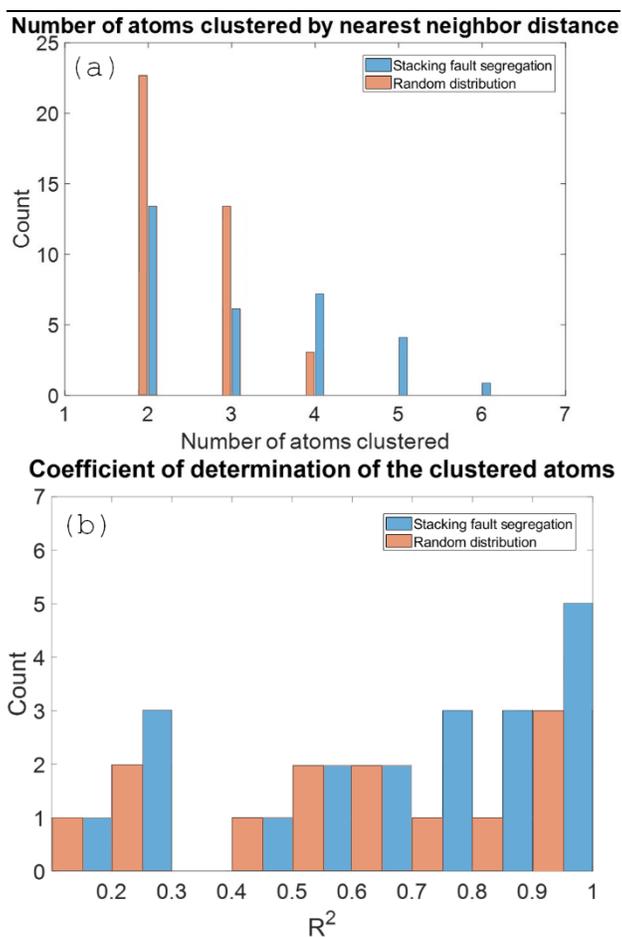

*Figure S11 - Histograms analyzing the clustered atoms from the simulations and the experimental results, and comparing it with a random distribution. (a) Shows the number of clustered atoms based on their size. (b) Histogram of the $R^2$ value of the polynomial fitting of the clusters.*

Table S1 - Parameters for the Ni-Ni interaction.

| β(r) | $c_0$ | $\lambda_0$ | $n_0$ | $c_1$ | $\lambda_1$ | $n_1$ |
|---|---|---|---|---|---|---|
| $dd\sigma$ | -15.1089 | 1.9168 | 0.7608 | -3.9348 | 0.5704 | 2.0557 |
| $dd\pi$ | 19.1848 | 1.7373 | 0.8815 | 5.8138 | 0.7475 | 2.0115 |
| $dd\delta$ | -9.5750 | 2.7407 | 1.2912 | -11.4674 | 1.9991 | 1.4293 |
| $E_{pair}$ | $c_{rep}$ | $\lambda_{rep}$ | $n_{rep}$ | | | |
| | 28.7918 | 0.7459 | 2.0701 | | | |
| $E_{emb}$ | $a_{emb}$ | $b_{emb}$ | | | | |
| | 2.6092 | 0.2064 | | | | |
| $E_{env}$ | B | $r_{core}$ | $\lambda_0$ | C | $\nu$ | m |
| | 3.6426 | 1.0 | 2.0 | 110.0 | 2.0 | 1.5 |
| Cut-offs | rcut | dcut | r2cut | d2cut | | |
| | 4.2 | 0.5 | 6.0 | 0.5 | | |

Table S2 - Parameters for the W-W interaction.

| β(r) | $c_0$ | $\lambda_0$ | $n_0$ | $c_1$ | $\lambda_1$ | $n_1$ |
|---|---|---|---|---|---|---|
| $dd\sigma$ | -22.8743 | 1.1202 | 0.8963 | 2.0460 | 0.2791 | 3.2146 |
| $dd\pi$ | 21.5035 | 1.5575 | 1.1492 | 1.2439 | 0.0847 | 2.4927 |
| $dd\delta$ | -1.3933 | 1.7957 | 1.1257 | -30.3070 | 1.6566 | 1.4821 |
| $E_{pair}$ | $c_{rep}$ | $\lambda_{rep}$ | $n_{rep}$ | | | |
| | 50.8147 | 0.4126 | 2.2251 | | | |
| $E_{emb}$ | $a_{emb}$ | $b_{emb}$ | | | | |
| | 3.3190 | 0.1918 | | | | |
| $E_{env}$ | B | $r_{core}$ | $\lambda_0$ | C | $\nu$ | m |
| | 1.6055 | 1.0 | 2.0 | 110.0 | 2.0 | 1.5 |
| Cut-offs | rcut | dcut | r2cut | d2cut | | |
| | 4.45 | 1.35 | 6.0 | 0.5 | | |

Table S3 - Parameters for the Ni-W interaction.

| $\beta(r)$ | $c_0$ | $\lambda_0$ | $n_0$ | $c_1$ | $\lambda_1$ | $n_1$ |
|---|---|---|---|---|---|---|
| dd$\sigma$ | -21.0610 | 1.3504 | 0.9835 | -3.8158 | 0.5934 | 1.8313 |
| dd$\pi$ | 17.6576 | 1.4972 | 0.9996 | 11.3672 | 1.0014 | 1.5544 |
| dd$\delta$ | -7.9148 | 1.7401 | 0.9999 | -16.4780 | 1.8550 | 1.3051 |
| $E_{pair}$ | $c_{rep}$ | $\lambda_{rep}$ | $n_{rep}$ | | | |
| | 22.4345 | 0.2393 | 2.8007 | | | |
| $E_{emb}$ | $a_{emb}$ | $b_{emb}$ | | | | |
| | 5.6630 | 0.3169 | | | | |
| $E_{env}$ | B | $r_{core}$ | $\lambda_0$ | C | $\nu$ | m |
| | 4.6799 | 1.0 | 2.0 | 110.0 | 2.0 | 1.5 |
| Cut-offs | rcut | dcut | r2cut | d2cut | | |
| | 4.2 | 0.5 | 6.0 | 0.5 | | |

$$M = \frac{L}{\xi R} \qquad (S1)$$

$$R = \frac{n \times d_{hkl}}{(1-\cos\theta)} \qquad (S2)$$

where L is the distance between the tip and detector screen, $\xi$ is the image compression factor (ICF), R is the radius of curvature of the tip, n is the number of terraces/rings, $d_{hkl}$ is the inter planar spacing, (hkl) is the Miller indice, and $\theta$ is the angle between two poles.

# References

Katnagallu, S., Stephenson, L. T., Mouton, I., Freysoldt, C., Subramanyam, A. P. A., Jenke, J., Ladines, A. N., Neumeier, S., Hammerschmidt, T., Drautz, R., Neugebauer, J., Vurpillot, F., Raabe, D. & Gault, B. (2019). Imaging individual solute atoms at crystalline imperfections in metals. *New Journal of Physics* **21**, 123020.

Natan, A. (2021). Fast 2D peak finder. https://www.mathworks.com/matlabcentral/fileexchange/37388-fast-2d-peak-finder (Accessed February 21, 2023).